\documentclass[twocolumn,runningheads]{svjour2}
\journalname{Astrophysics and Space Science}
\smartqed  
\usepackage[authoryear,round]{natbib}
\usepackage{graphicx}
\usepackage{psfig}
\begin{document}

\title{Chandra Smells a RRAT}

\subtitle{X-ray Detection of a Rotating Radio Transient}


\author{Bryan M.\ Gaensler \and
Maura McLaughlin \and
Stephen Reynolds \and
Kazik Borkowski \and
Nanda Rea \and
Andrea Possenti \and
Gianluca Israel \and
Marta Burgay \and
Fernando Camilo \and
Shami Chatterjee \and
Michael Kramer \and
Andrew Lyne \and
Ingrid Stairs}

\authorrunning{Gaensler et al.} 

\institute{B. M. Gaensler and S. Chatterjee \at
              Harvard-Smithsonian Center for Astrophysics, Cambridge MA, USA \\
              \email{bgaensler@cfa.harvard.edu} \\
              Present address: The University of Sydney, NSW, Australia
          \and
          M. McLaughlin \at University of West Virginia \and
          S. Reynolds and K. Borkowski \at North Carolina State University \and
          N. Rea \at SRON Netherlands Institute for Space Research \and
          A. Possenti and M. Burgay 
\at Osservatorio Astronomico di Cagliari \and
          G. Israel \at Osservatorio Astronomico di Roma \and
          F. Camilo \at Columbia University \and
          M. Kramer and A. Lyne \at Jodrell Bank Observatory \and
          I. Stairs \at University of British Columbia
}

\date{Received: date / Accepted: date}

\maketitle

\begin{abstract}
``Rotating RAdio Transients'' (RRATs) 
are a newly discovered astronomical phenomenon,
characterised by occasional brief radio bursts, with average intervals
between bursts ranging from minutes to hours.  The
burst spacings allow identification of periodicities, which fall in the
range 0.4 to 7 seconds.  The RRATs thus seem to be rotating neutron stars,
albeit with properties very different from the rest of the population.
We here present the serendipitous detection with the {\em Chandra X-ray
Observatory}\ of a bright point-like X-ray source coincident with one of
the RRATs. We discuss the temporal and spectral properties of this X-ray
emission, consider counterparts in other wavebands, and interpret these
results in the context of possible explanations for the RRAT population.

\end{abstract}

\section{Introduction}
\label{intro}

Astronomers are still coming to terms with the fact that the zoo of isolated
neutron stars
harbours an increasingly diverse population. 
Objects of considerable interest include
radio pulsars,
anomalous
X-ray pulsars, soft gamma repeaters, central compact objects
in supernova remnants (SNRs), and
dim isolated neutron stars. A startling new discovery has now
forced us to further expand this group: \cite{mll+06} have
recently reported the detection
of eleven ``Rotation RAdio Transients'', or ``RRATs'', characterised
by repeated, irregular radio bursts, with burst
durations of 2--30~ms, and intervals between bursts of $\sim4$~min to
$\sim3$~hr. The RRATs are concentrated at low Galactic latitudes, with
distances implied by their dispersion measures of $\sim2-7$~kpc.

For ten of the eleven RRATs discovered by \cite{mll+06}, an analysis of the
spacings between repeat bursts reveals an underlying spin period, $P$,
and also in three cases, a spin period derivative, $\dot{P}$. The observed
periods fall in the range 0.4~s~$<P<$~7~s, which generally overlap with
those seen for the radio pulsar population.

For the three RRATs with values measured for both $P$ and $\dot{P}$,
a characteristic age, $\tau_c$, and a dipole surface magnetic
field, $B$, can both be inferred, as listed in Table~\ref{tab_rrats}.
These sources can also
be placed on the standard ``$P-\dot{P}$ diagram'', and can thus
be compared to other populations of rotating neutron star:
\begin{itemize}
\item RRAT~J1317--5759
has properties typical of radio pulsars,
except for a relatively long spin period.
\item RRAT~J1819--1458
is very young, with a high magnetic field.
On the $P-\dot{P}$ diagram, it is located in the upper-right region,
in the same area occupied by the magnetars and by the high-field
radio pulsars.
\item RRAT~J1913+3333
has spin properties indistinguishable
from the bulk of radio pulsars.
\end{itemize}

We here present the X-ray detection of RRAT~J1819--1458, which provides
a further point of comparison with the various neutron star population.
The details of this detection and its interpretation are discussed by
\cite{rbg+06}.

\begin{table}[hbt]
\caption{Properties of the three RRATs with measured period derivatives.}
\centering
\label{tab_rrats}       
\begin{tabular}{lccc}
\hline\noalign{\smallskip}
Source & $P$ (sec) & $\tau_c$ (Myr) & $B$ ($10^{12}$~G)  \\[3pt]
\tableheadseprule\noalign{\smallskip}
RRAT~J1317--5759 & 2.6 & 3.3 & 5.83 \\
RRAT~J1819--1458 & 4.3 & 0.12 & 50 \\
RRAT~J1913+3333 & 0.92 & 1.9 & 2.7 \\
\noalign{\smallskip}\hline
\end{tabular}
\end{table}

\section{X-ray Emission from RRAT~J1819--1458}

\subsection{Detection}

As mentioned above, RRAT~J1819--1458 is very young, and has a high
surface magnetic field. This source bursts every $\sim3$~min, making
it the most active of the known RRATs.  Its dispersion measure of
$196\pm3$~pc~cm$^{-3}$ corresponds to an estimated distance of 3.6~kpc,
with considerable uncertainty.

\begin{figure}[bht]
\centerline{\psfig{file=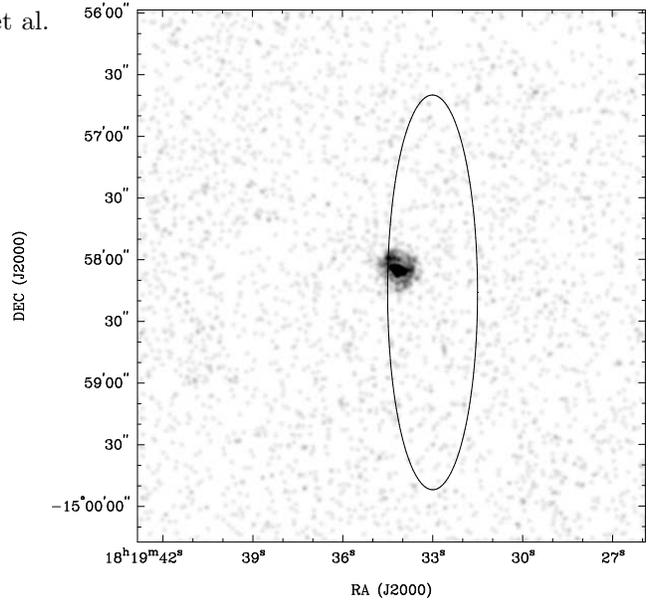,width=0.48\textwidth,angle=270}}
\caption{A {\em Chandra}\ ACIS image of RRAT~J1819--1458, observed
$11'$ off-axis, and smoothed with a gaussian of FWHM $2''$. 
The ellipse shows the $3-\sigma$ error ellipse for the position
of the RRAT, as derived by \cite{mll+06} from radio timing.}
\label{fig_image}
\end{figure}

RRAT~J1819--1458 sits only
$\sim11'$ from the young
SNR G15.9+0.2.  This source was the target of a
30~ks observation with {\em Chandra}\ ACIS in May~2005, and
RRAT~J1819--1458 fortuitously falls within the field of view
\citep{rbg+06}.  As shown in Figure~\ref{fig_image}, there is a clear detection
of a bright, unresolved X-ray source within the error ellipse for
RRAT~J1819--1458.  Using the X-ray
differential source count distribution for the Galactic plane of
\cite{smk+01}, we
find that the probability of finding a field source of this count
rate at this position is $<10^{-4}$, and conclude that we have both
identified and localised the X-ray counterpart of RRAT~J1819--1458.

\subsection{Spectrum and Variability}

\begin{figure}[bht]
\centerline{\psfig{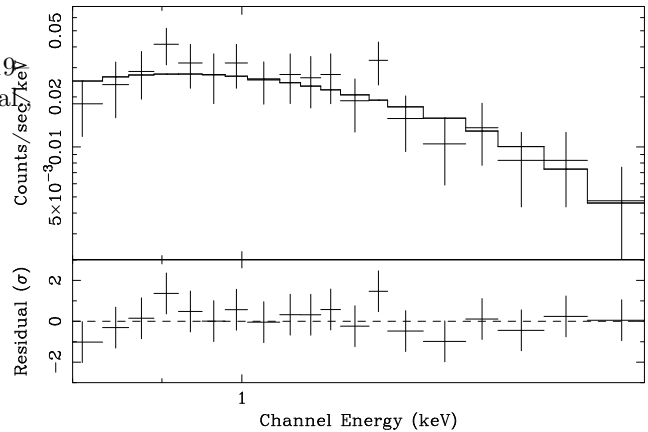}}
\caption{The X-ray spectrum of RRAT~J1819--1458, as observed
with {\em Chandra}\ ACIS \citep{rbg+06}. The data points show
the measurements, while the solid line shows the best-fit absorbed
blackbody described in the text.} 
\label{fig_spec}
\end{figure}

We have extracted $524\pm24$ counts from the X-ray counterpart
to RRAT~J1819--1458, the spectrum from which
is shown in Figure~\ref{fig_spec}. While this is insufficient for detailed
spectral modelling, it still yields crucial information.

Fitting simple absorbed power-law and blackbody models, we find
that the spectrum is a poor fit to the former, but a good fit to
the latter \citep{rbg+06}. 
At a distance of $3.6d_{3.6}$~kpc, the inferred blackbody
radius (as viewed at infinity) is $20d_{3.6}$~km, which is consistent
with standard neutron star equations of state, given the uncertainty in the
distance estimate.  For this blackbody fit, the foreground absorbing
column is $N_H = 7^{+7}_{-4} \times 10^{21}$~cm$^{-2}$, and the
surface temperature (at infinity) is $kT_\infty = 120\pm40$~eV.  In
the energy range 0.5--8.0~keV, the unabsorbed flux is
$2\times10^{-12}$~ergs~cm$^{-2}$~s$^{-1}$ and the isotropic luminosity
is $3.6d_{3.6}^2 \times 10^{33}$~ergs~s$^{-1}$.

Given the extreme variability of the source seen at radio wavelengths,
it is of interest to quantify the level of X-ray variability seen
in this observation.
Examination of individual CCD frames (at a time-resolution of
3.2-seconds) shows no evidence for individual brief X-ray
bursts that might be a counterpart to the RRAT phenomenon,
with a 3-$\sigma$ upper limit on the observed fluence of any
burst of $3 \times 10^{-11}$~ergs~cm$^{-2}$ in the 0.5--8~keV
energy range.
The data also show no evidence of variability
on any time-scale ranging from 3.2~s
to 5 days (the time-span covered
by the observations). Although the time resolution is only
slightly shorter than the spin period, $P = 4.3$~s, we do have
limited sensitivity to an aliased pulsed signal.
However, we find no pulsations down to a 3-$\sigma$ pulsed fraction limit of
70\% for a sinusoidal pulse profile \citep{rbg+06}.

\subsection{Data At Other Wavelengths}

With an accurate position for RRAT~J1819--1458 in hand from our {\em
Chandra}\ position, we have searched for a counterpart to this source
at other wavelengths, using archival VLA, 2MASS and GLIMPSE data.
No detection is made in any of these data.

\begin{figure}
\centerline{\psfig{file=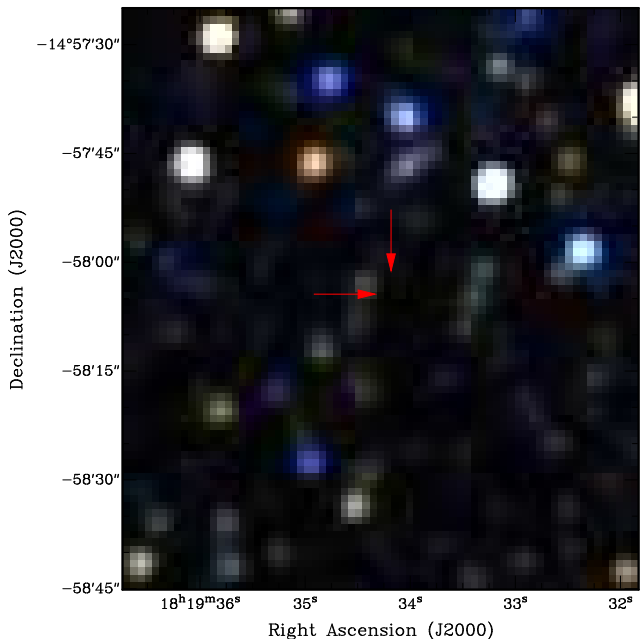,width=0.48\textwidth}}
\caption{A 2MASS image of the field surrounding RRAT~J1819--1458,
with $J$, $H$ and $K$ bands shown in blue, green and red, respectively.
The arrows mark the position of the RRAT, as derived
from {\em Chandra}\ observations. The upper limits on the
infrared emission from RRAT~J1819--1458 are $J > 15.6$,
$H > 15.0$ and $K > 14.0$.}
\label{fig_2mass}
\end{figure}

The non-detection in 2MASS data, shown in Figure~\ref{fig_2mass},
allows us to put a lower limit on the X-ray to infrared flux ratio
of $f_X/f_{IR} > 0.7$, which rules out most stellar counterparts
\citep[see][]{kfg+04}.
Furthermore, the foreground column density inferred from the X-ray
spectral fit is substantially smaller than the integrated column
through the Galaxy at this position, arguing against the X-ray
emission being from a background galaxy or cluster. These arguments,
combined with the good blackbody fit to the X-ray spectrum, imply that
the most reasonable interpretation for the X-ray emission is thermal
emission from a neutron star surface.

\section{Comparison with Other Sources}
\label{sec_compare}

It is unclear whether the RRATs are a completely new group of neutron
stars, or are a new manifestation of one of the previously known
classes of object. Since all classes of young neutron star are X-ray
emitters, the detection reported here provides vital new information
which allows us to begin to discriminate between various interpretations
for the RRATs.

We first note that while many neutron stars are pulsed in X-rays,
those which produce thermal emission inevitably exhibit low pulsed
fractions, due to gravitational bending of the light emanating from
their surfaces \citep[e.g.,][]{pod00}. 
Thus our relatively poor limit on pulsed fraction
is unsurprising and unconstraining.

Despite its location near the magnetars on the $P-\dot{P}$
diagram, the X-ray properties of RRAT~J1819--1458 are distinct from
those seen for magnetars: the RRAT is much colder and less luminous
than the magnetars, and 
apparently lacks the hard X-ray tail seen for these
sources, at our current level of
sensitivity.  Furthermore, the magnetar birth rate is well below
that estimated for the RRAT population \citep{ptp06}.
The only possible link to magnetars is with the transient magnetar
XTE~J1810--197 which, while in quiescence,
had a surface temperature $kT_\infty \approx
150-180$~eV \citep{ims+04,ghbb04}, comparable to that seen here.

The X-ray temperature of RRAT~J1819--1458 is also well below
those of the ``central compact objects'', while its temperature
and luminosity are above those of most of
the dim isolated neutron stars.

The one population whose X-ray properties provide a reasonable match
to that of the RRAT is that of radio pulsars, for which sources
with ages around 100~kyr show spectra very similar to what we have
found here. For example, PSR~J0538+2817 is 30~kyr old and has
$kT_\infty = 160$~eV, while PSR~B0656+14 is 110~kyr old and has
$kT_\infty = 70$~eV \citep[see][for details]{rbg+06}. The
X-ray emission from RRAT~J1819--1458 thus suggests that this source
is a normal radio pulsar, albeit one that produces
unusual radio bursts. Additional evidence to support
this possibility is the subsequent discovery from
PSR~B0656+14 of RRAT-like behaviour \citep{wsrw06}. If PSR~B0656+14
was placed at the distance of RRAT~J1819--1458, its 
occasional bright radio bursts would still be seen, but
not the underlying regular train of pulsations.

One point to note is that the inferred surface magnetic field
strength of RRAT~J1819--1458 is more than an order of magnitude
greater than those of PSRs~J0538+2817 and B0656+14 discussed above.
Two radio pulsars with comparable magnetic fields that have been
detected in X-rays are PSRs~J1718--3718 \citep{km05} and J1119--6127
\citep{gkc+05}. These sources  show temperatures ($kT \sim 150-200$~eV) and
luminosities ($\sim10^{32}-10^{33}$~ergs~s$^{-1}$) comparable to that
of RRAT~J1819--1458, although both sources are probably much younger
(35 and 1.7~kyr, respectively) and, in contrast to RRAT~J1819--1456,
have X-ray luminosities less than their spin-down luminosities.

\section{Possible Interpretations}

If the RRATs are normal radio pulsars as proposed above, we then
need to explain their transient behaviour.  One possibility, discussed
by \cite{zgd06}, is that RRATs are pulsars that are no longer active,
but for which a temporary ``star spot'' with multipole field components
emerges above the surface. This magnetic field component could
temporarily reactivate the radio beaming mechanism, producing the
observed bursts. 

The difficulty with this possibility is that none of the RRATs with
known values of $\dot{P}$ appear to be near the ``death line'' in
the $P-\dot{P}$ diagram, beyond which the pulsar mechanism is
expected to turn off.  To account for this, \cite{zgd06} have proposed that
RRATs have dipole fields offset from their centres, causing their
magnetic fields to be over-estimated.  However, the X-ray temperature
seen for RRAT~J1819--1458 is consistent with it being a $\sim100$-kyr
old neutron star (see Sec.~\ref{sec_compare} above), 
in reasonable agreement with
its characteristic age as listed in Table~\ref{tab_rrats}, and
confirming that if this is a radio pulsar, it is as yet nowhere near death.

\cite{zgd06} also consider the possibility that RRATs are caused by a brief
reversal of the direction in which radio beams are emitted. They
draw upon the previous work of \cite{gjk+94}, who showed that whenever
PSR~B1822--09 produces ``nulls'' in its main pulse, it produces a
much weaker interpulse, which has been interpreted by
\cite{dzg05} to correspond
to radiation temporarily emitted in the opposite direction. If one
invokes an emitting geometry in which the main pulse is never seen,
then the interpulse alone, appearing only when the unseen main pulse nulls,
might correspond to RRAT-like emission. The difficulty with this
interpretation is that in PSR~B1822--09 this reversal lasts for
several minutes, in contrast to the RRATs, for which multiple
bursts in succession are yet to be observed.

Finally, the RRAT mechanism might be produced by interaction of
the neutron star with an equatorial fallback disk or with orbiting
circumpolar debris.  Accretion from a disk should usually quench the
radio emission mechanism, but sporadic drops in the accretion rate could
allow the radio beam to turn on for a fraction of a second, producing
the RRAT phenomenon \citep{li06}.  However, this possibility is at odds
with the behaviour seen by \cite{wsrw06} for PSR~B0656+14, in which the
RRAT-like bursts are superimposed on an underlying persistent series of
faint radio pulsations.  Alternatively, episodic injection of material
from a circumpolar asteroid belt could temporarily activate a quiescent
region of the magnetosphere, producing a RRAT burst, in some cases from
a radio pulsar that is normally beaming away from us \citep{cs06}.

\section{Future Observations and Conclusions}

A variety of forthcoming observations should be able to cast more
light on the possibilities discussed above, and on the spatial and spin
distributions of these sources. Several groups have already undertaken new
radio searches for RRAT-like emission from other classes of neutron stars.
Near-infrared observations with the VLT have also been obtained for some
RRATs, to provide stronger constraints on the X-ray to infrared flux
ratio of these sources, and to look for hints of a fossil / fallback disk
\citep[see][]{wck06}.
Meanwhile, in a project called ``Astro Pulse'', the data from the Arecibo
SETI@Home project are being reanalysed for short transient signals. With
the sensitivity of Arecibo, many RRATs may be found in this analysis.

Finally, following on from our X-ray detection discussed here, {\em
XMM-Newton}\ observations of two RRATs have been approved for Cycle~5:
of RRAT~J1819--1458 (to obtain a better spectrum and to properly search
for pulsations) and of RRAT~J1317--5759 (to see if it too emits detectable
X-rays).

A final point to make is that it is already clear from the 11 RRATs
known that the birth rate of these objects is $\sim3-4$ times that
previously estimated for all radio pulsars \citep{mll+06,ptp06}. We are thus
only seeing the upper tip of a large transient distribution.
When one combines this with the populations of nullers
\citep{bac70}, ``bursters'' \citep[e.g., PSR~J1752+2359;][]{lwf+04}, 
``winkers'' \citep[PSR~B1931+24;][]{klo+06}
and ``burpers'' \citep[GCRT J1745-3009;][]{hlk+05}, 
it becomes clear that a full study of
the transient radio sky is needed \citep{clm04}.
The next generation of radio telescopes, with the capability of
monitoring very wide fields of view (e.g., LOFAR, xNTD and the SKA) should
make many further discoveries of such phenomena.

\begin{acknowledgements}
B.M.G. acknowledges the support of NASA through LTSA grant
NAG5-13023 and of an an Alfred P.\ Sloan Fellowship. 
\end{acknowledgements}


\bibliographystyle{abbrvnat}
\bibliography{journals,modrefs,psrrefs,crossrefs}

\end{document}